\documentclass[fleqn,usenatbib]{mnras}
\usepackage{newtxtext,newtxmath}

\usepackage[T1]{fontenc}
\usepackage{ae,aecompl}

\usepackage{graphicx}	

\def\msun{ {\rm M_\odot}}

\newcommand{\kpc}{{\rm kpc}}
\newcommand{\vlos}{v_{\rm los}}

\newcommand{\kms}{ {\rm km~s\textsuperscript{-1}}}

\newcommand{\oversim}[2]{\protect{\mbox{\lower0.5ex\vbox{%
   \baselineskip=0pt\lineskip=0.2ex
   \ialign{$\mathsurround=0pt #1\hfil##\hfil$\crcr#2\crcr\sim\crcr}}}}} 
\catcode`\"=\active\let"=\"  

\def\3{{\ss} }

\def\c12{{1\over 2}}

\def\plusplus{\raise 0.3ex\hbox{${\scriptstyle ++}$}{}}

\def\and{{{\rm M}31}}

\begin{document}
\label{firstpage}
\pagerange{\pageref{firstpage}--\pageref{lastpage}}

\title[The LMC in the North]
 {Tidally stripped halo stars from the Large Magellanic Cloud in the Galactic North}

\author[Petersen, Pe{\~n}arrubia, Jones] 
{Michael~S.~Petersen,$^1$\thanks{michael.petersen@roe.ac.uk}
  Jorge~Pe{\~n}arrubia$^{1,2}$, Ella Jones$^1$ \\ 
$^1$Institute for Astronomy, University of Edinburgh, Royal Observatory, Blackford Hill, Edinburgh EH9 3HJ, UK\\ $^2$Centre for Statistics, University of Edinburgh, School of Mathematics, Edinburgh EH9 3FD, UK
}

\maketitle

\begin{abstract}
We examine whether the Large Magellanic Cloud (LMC) is currently losing its stellar halo to Milky Way (MW) tides. We present a live $N$-body model for the ongoing MW-LMC interaction that predicts a prominent stream of stars tidally stripped from the progenitor LMC. We use this model to define a strategy to search for stripped material in kinematic space. Of the available stellar tracers, we conclude that samples of RR Lyrae stars provide the highest density of kinematic tracers at present. Using a sample of RR Lyrae stars with Gaia EDR3 astrometry we show that the LMC stellar halo in the Southern Galactic hemisphere extends at least out to $\sim 30^\circ$ from the galaxy centre. In addition, several leading arm candidates are found in the Northern hemisphere as far above the disc plane as \(b=+34^\circ\) (at 68\(^\circ\) from the LMC).
\end{abstract} 

\begin{keywords} galaxies: Galaxy: halo---galaxies: haloes---galaxies: kinematics and dynamics---galaxies: evolution---galaxies: structure \end{keywords}

\section{Introduction} \label{sec:introduction}

Our Milky Way (MW) galaxy halo is actively undergoing dynamical evolution owing to the recent infall of the Large Magellanic Cloud (LMC). A number of effects have recently been observed, including the reflex motion of the MW disc in response to the LMC \citep{petersenpenarrubia20b,erkal21,vasiliev21} and the purported wake in the MW stellar halo \citep{garavitocamargo19,belokurov19,conroy21}. With an estimated mass of \(1.4\times10^9\msun\) \citep{deason19}, the stellar halo of the MW is an excellent tracer of the past history and current dynamical state of the MW. However, the MW halo evolution pales in comparison to the expected evolution of the LMC in the tidal field of the MW. Recent models predict significant LMC mass loss and associated features \citep{garavitocamargo21}. Some observations possibly indicative of this mass loss have been detected. For example: a large amount of gas coincident with the inferred LMC trajectory \citep{nidever08,nidever10}, stars in the trailing arm of the LMC \citep{zaritsky20}, and a young star cluster in the leading arm of the LMC \citep{pricewhelan19,nidever19b}.

Studies of the LMC have found evidence for non-equilibrium features, such as a perturbed outer stellar disc \citep{mackey16,belokurov19b,cullinane20}  and apparent stream-like substructures \citep{belokurov16,navarrete19}. The LMC also appears to have an extended stellar `envelope' \citep{majewski09,nidever19a}. The stellar envelope (beyond~\(\approx13\) kpc, or $\simeq 15^\circ$, from the LMC centre) is measured to have a relatively shallow power-law slope, with \(\rho_{\rm envelope}\propto r^{-2.2}\) continuing to~\(\approx20\)~kpc from the LMC centre \citep{nidever19a}. In a hierarchical assembly paradigm, galaxies with the mass of the LMC are expected to host an extended stellar halo mainly populated with old stars. A clear theoretical prediction on the luminosity and spatial distribution of halo stars is complicated by the fact that the LMC is being perturbed by the Milky Way. 
Unfortunately,
many models that follow the MW-LMC interaction do not include a responsive (i.e. live) LMC, and are therefore ill-suited to predict the kinematics of stars and dark matter at large radii from the LMC centre. Yet, a detection of the LMC halo would open up promising avenues for studying a record of the assembly of the LMC and help place the MW-LMC system in the context of cosmological evolution. In this regard, the diversity of stellar halos around nearby galaxies suggests that the LMC could plausibly have a stellar halo as low as a hundredth of the stellar content or as high as a tenth of the stellar mass \citep{bell17,smercina20}. Adopting the stellar mass of \(2.7\times10^9\msun\) measured by \citet{vandermarel02} suggests that the LMC may host a halo with stellar mass greater than \(3\times10^7\msun\).

An unambiguous detection of the LMC stellar halo is also complicated by the poor understanding of the dynamics of the MW at the distance of the LMC ($\gtrsim 50$ kpc), with current samples of tracers with full kinematic information only numbering in the hundreds at such distances \citep{petersenpenarrubia20b}. Fortunately, the Gaia mission \citep{gaiamission} has produced an incredible wealth of astrometric data, as well as auxiliary classifications for unique types of stars. In particular, the RR Lyrae sample(s) of Gaia DR2 provide an all-sky\footnote{Albeit coverage-biased, as discussed below.} look at the distribution of proper motions \citep{holl18,clementini19}. RR Lyrae stars are old, metal-poor, and bright standard candles -- unlocking an extra data dimension -- that may be used to trace the distribution of distant stars. RR Lyrae stars trace the stellar halo of the MW out to large distances \citep{petersenpenarrubia20b}. These stars are present in large numbers in the Sagittarius dwarf and its associated tidal stream \citep{ramos20}, as well as in both globular clusters and satellite galaxies. We are therefore motivated to make predictions for the location of the extended LMC stellar halo using RR Lyrae as kinematic tracers to test whether the LMC entered the MW embedded in an extended stellar halo.

In this paper, we present a live $N$-body model that describes the MW-LMC pair in order to construct mock observations, guide our strategy to search LMC halo stars, and find promising candidates for spectroscopic follow-up that establishes LMC membership.

\section{Live $N$-body model} \label{sec:models}

\subsection{Initial conditions} \label{subsec:initialconditions}

We construct a live three-component model, consisting of an initially spherical LMC (\(N=5\times 10^6\)), an initially spherical MW halo (\(N=10^7\)), and a stellar disc for the MW (\(N=10^6\)). We first realise the individual components of the model in virial units, and then scale the model to match  the observed rotation curve constraints and estimated overall mass of the MW-LMC system. The LMC model is a spherically-symmetric Navarro-Frank-White (NFW) dark matter halo radial profile \citep{navarro97} given by $\rho_{\rm NFW}(r) \propto r_s^3r^{-1}\left(r+r_s\right)^{-2}$. The scale radius of the LMC is set to be $r_s=0.06R_{\rm vir,LMC}$, where $R_{\rm vir,LMC}$ is the LMC virial radius. We apply an error function truncation such that the initial halo profile is $\rho_h(r)=\frac{1}{2}\rho_{\rm NFW}(r)\left(1-{\rm erf}\left[(r-r_{\rm trunc})/w_{\rm trunc}\right]\right)$. The truncation parameters are $r_{\rm trunc}=2R_{\rm vir,LMC}$ and $w_{\rm trunc}=0.3R_{\rm vir,LMC}$. We realise the spherical halo by Eddington inversion \citep{binney08}. We verify that the LMC is in equilibrium when run in isolation.

The total MW model consists of a stellar disc embedded in a dark matter halo. We realise a spherical MW halo following a basic spherically-symmetric MW model. As for the LMC, we choose an NFW profile for the dark matter halo, changing the \({}_{\rm vir,LMC}\) subscripts to \({}_{\rm vir,MW}\). Owing to the large uncertainties in the outer halo of the MW, we simply choose a fiducial model with a scale radius of \(r_s=0.056R_{\rm vir}\). The truncation parameters are the same as for the LMC. The stellar disc density is given by \(\rho_d(r,z)~=~(M_{\rm d}/8\pi z_0R_d^2)~e^{-r/R_d} {\rm sech}^2~(z/z_0)\) where \(M_d\) is the disc mass, \(R_d=0.01R_{\rm vir}\) is the disc scale length, and \(z_0=0.002R_{\rm vir}\) is the disc scale height. We choose \(M_d=0.05M_{\rm vir,MW}\). We select the initial positions in the disc via an acceptance--rejection algorithm. We select the velocities by solving the Jeans equations in the disc plane. We refer the reader to \citet{petersen21} for details. We take a 20\% LMC:MW mass ratio inspired by \citet{penarrubia16} as our model case, such that \(M_{\rm vir,LMC}=0.2M_{\rm vir,MW}\). The relationship of the virial radii is chosen to be \(R_{\rm vir,LMC}=0.5R_{\rm vir,MW}\).

We integrate the simulations using the basis function expansion \(N\)-body code {\sc exp} \citep{weinberg99,petersen21b}. The relative positions of the MW-LMC pair is set by finding a trajectory that nearly satisfies the observed constraints for the centre of the present-day LMC. We call this time $T=0$. To analyse the match to the MW-LMC system, we scale the simulation such that \(R_d=3\)kpc, \(z_0=600\) pc, and \(v_c(R=R_\odot)=229\kms\) , in line with typical values for the MW \citep{blandhawthorn16}. For the chosen parameters, this results in a mass \(M_{\rm vir,MW}(R<300{\rm kpc})=10^{12}M_\odot\). The mass of the LMC immediately follows owing to the choices defined above, \(M_{\rm vir,LMC}(R<150{\rm kpc})=2\times10^{11}M_\odot\). Using the initial MW disc and halo potential, the LMC trajectory is rewound in time to $T=-4$ Gyr,  at which time the LMC is \(\approx600\) kpc from the MW centre. We neglect the effect of the LMC on the MW during the rewind. When the model is evolved forwards in time, the centre of the MW responds to the infall of the LMC, changing the apparent trajectory. To resolve this difference, we iteratively search the nearby parameter space of the MW-LMC initial conditions using low-resolution versions of the model. For a given trajectory we first measure the displacement of the disc. We then re-calculate the analytic trajectory using the distance the MW disc travels and repeat the process until a tolerance value is reached. We use the LMC centre and mean proper motion for the LMC computed from the EDR3 data for the LMC \citep{gaia_lmc21} as well as the distance from \citet{pietrzynski19} and the line-of-sight velocity from \citep{vandermarel02}: \(\left(d,\alpha_{\rm LMC},\delta_{\rm LMC}\right)=\left(49.59~{\rm kpc}, 81.28^\circ,-69.78^\circ\right)\) and \(\left(v_{\rm los},\mu_{\alpha^\star,{\rm LMC}},\mu_{\delta,{\rm LMC}}\right)=\left(262.2~\kms,1.7608 ~{\rm mas/yr},0.308~{\rm mas/yr}\right)\). In the particular case of the model presented here, we find that we match the observed location on the sky (within 2 degrees), distance (within 2 kpc), \(\mu_l\) and \(\mu_b\) (both within 0.1 mas/yr) of the LMC, but the line-of-sight velocity is modestly different ($180~\kms$ in the model versus $262.2~\kms$ observed). This likely stems from the particular choice of MW and LMC mass distributions; future work is needed to search through model {\it and} trajectory space to find the span of models that can accurately match the observables. Despite this mismatch, our model is sufficiently close to the current data for the LMC trajectory so as to inform a search for the extended LMC stellar halo (see Section~\ref{subsec:modelfindings}). 

\begin{figure} \centering \includegraphics[width=3.4in]{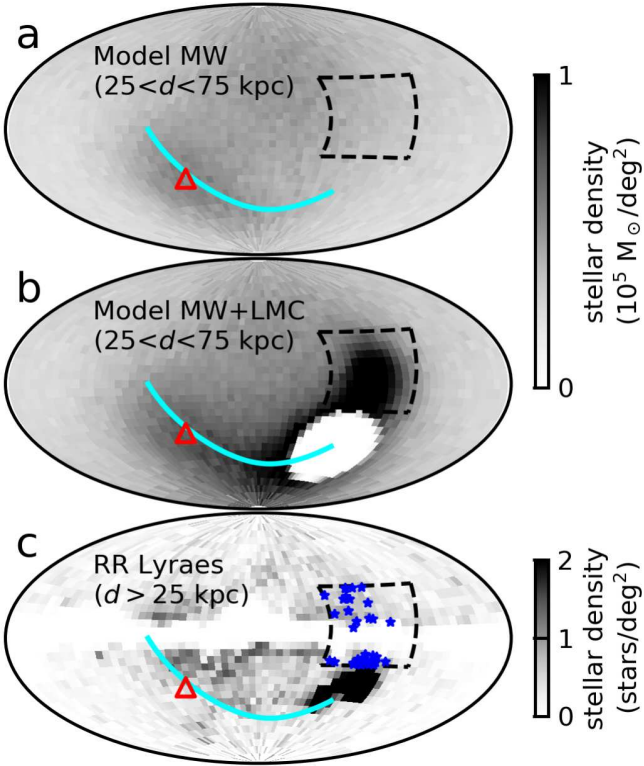} 
\caption{\label{fig:skydensity} Comparison of model and data on-sky source densities, shown in Aitoff-projected galactic coordinates \((\ell,b)\). In our projection, \(\ell\) increases to the left. In each panel, we bin the source density in \(2^\circ\) patches of the sky, shown in greyscale (panels a and b have one normalisation, panel c another, as indicated). Panel a: data for the stellar density in the \(\rho_{\rm MW,\star}\propto r^{-3}\) model MW (i.e. no LMC) only, the range 25$<d_\odot<$75 kpc. Panel b: data for the combined MW+LMC  \(\rho_{\rm \{MW,LMC\},\star}\propto r^{-3}\) model, demonstrating the significant amount of LMC stellar material in the defined search region with the given model assumptions. We eliminate a \(25^\circ\) region around the LMC to improve the visibility of lower surface density structures. Panel c: the RR Lyrae dataset from the combined Gaia DR2 sample (Sec.~\ref{subsec:sample}). RR Lyrae stars passing the kinematic selection criteria described in the text are shown as blue stars. In each panel, we outline the `search region' (Sec.~\protect{\ref{subsec:modelfindings}}) with dashed black curve, show the model trajectory of the LMC over the past 2 Gyr as a cyan curve, and show the location of the data-estimated `apex' \protect{\citep[the direction of travel of the MW disc,][]{petersenpenarrubia20b}} as a red triangle.} \end{figure}

\subsection{Mock data set construction}\label{subsec:construction}

We create two data products from the simulations. The first is a {\it model} for the stellar halo that follows a specified mass density distribution. The second is a {\it mock} dataset for the observed superposition of the stellar MW and LMC halos, matched to the RR Lyrae dataset described in Section~\ref{subsec:sample}. When the entire model is analysed, we refer to it as the model; when the matched distribution is analysed, we refer to it as the mock.

To estimate the population of the stellar halo in the MW and LMC, we re-sample the dark matter distribution functions for the MW and LMC, following the distribution function re-weighting procedure described in \citet{errani20}. We tested different profiles for the stellar halos, settling on a simple power law with \(\rho_{\rm \{MW,LMC\},\star}\propto r^{-3}\) after finding that other halo profiles only changed the model findings in amplitude, rather than phenomenology. After computing relative weights of particles, we resample the weighted particles using acceptance-rejection to construct a `complete' stellar halo catalog for the MW and LMC (each comprising approximately 50\% of the number of original particles). We set the mass of the MW stellar halo to be \(1.4\times10^9\msun\) and the mass of the LMC stellar halo to be \(2.8\times10^8\msun\).

To `observe' the simulation, we transform to local standard of rest observations with \(\vec{x}_{\rm obs} = \vec{x}_{\rm disc frame}+\vec{x}_\odot\) and \(\vec{v}_{\rm obs} = \vec{v}_{\rm disc frame}+\vec{v}_\odot\) where \((\vec{x},\vec{v})_{\rm disc frame}\) are the quantities in the rest frame of the disc. We define the transformation between the heliocentric and galactocentric frames by placing the sun (in a standard left-handed coordinate system) at \(\vec{x}_\odot = (x,y,z)=(8.17,0.0,0.02)\) kpc \citep{gravity19,bennett19}, with velocity  \(\vec{v}_\odot = (u,v,w)=(-12.9,245.6,7.78)\) km s\(^{-1}\) \citep{drimmel18}. We then convert the Cartesian coordinates to observed spherical coordinates in the heliocentric frame, \((x,y,z,u,v,w)\to(d,\ell,b,v_{\rm los},v_\ell,v_b)\). We also define an auxiliary Magellanic Stream (MS) heliocentric rotated coordinate system \((d,\phi_1,\phi_2)^{\rm LMC}\) where the LMC is placed at the \((\phi_1,\phi_2)=(0^\circ,0^\circ)\) origin, as defined for the gaseous MS by \citet{nidever08}. 

In panels a and b of Figure~\ref{fig:skydensity}, we show the primary on-sky density features of the model: two overdensities in the MW stellar halo, indicative of the reflex motion of the MW disc in response to the presence of the LMC and the deformation of the MW stellar halo (see further discussion in Sec.~\ref{subsec:lmcfeatures}), and an extended LMC stellar halo stretched along the MS. Panel b of Figure~\ref{fig:skydensity} shows the combined MW and LMC material in a thick shell from \(25<d<75\) kpc. The greyscale indicates integrated on-sky mass density. To facilitate seeing the relatively low-contrast LMC stellar halo, in panel b we remove all stars within 25\(^\circ\) of the model LMC centre. By mass, the LMC material dominates the thick shell. Owing to the interplay of multiple global effects, it is clear that a density distribution on the sky will be at best ambiguous to interpret, and is highly model-dependent.

Motivated by our model, we focus on predictions for the leading arm of the stellar halo of the LMC at relatively large distances from the LMC centre\footnote{One could also use the model the study the trailing arm of the LMC. However, the trailing arm is at larger distances, so we focus on the leading arm.}. To construct mock data sets we use the sample described in Section~\ref{subsec:sample}. We use acceptance-rejection of the stellar halo model to build a mock catalogue that reproduces the heliocentric distance cumulative curve and overall number of stars in the observed data set. We do not attempt to model the on-sky distribution. That is, we do not tune the number of LMC stars beyond the choices inherent in making the model, and we do not mask any extinction regions or known substructure locations.

\subsection{Model and Mock Data Findings} \label{subsec:modelfindings}

To validate our mock model against observational data, we choose two 6-dimensional observed signposts and check the validity of the model again. First, the location and kinematics of the LMC itself, defined in Section~\ref{subsec:initialconditions}. Second, the star cluster Price-Whelan~1 (PW1, \(\ell=288.8^\circ,~b=32.0^\circ\)), thought to be in the leading arm of the MS \citep{pricewhelan19,nidever19b}. We update the proper motion kinematics for PW1 by cross-matching the \citet{nidever19b} sample with Gaia EDR3, finding that 26 of the 28 stars with line-of-sight velocities have valid EDR3 results consistent with cluster membership\footnote{Using names from \protect{\citet{nidever19b}}: PW-11 has {\tt ruwe}\(\sim\)2, indicative of uncertain Gaia proper motion measurements, and PW-26 now has proper motions inconsistent with the rest of the cluster.}. PW1 lies squarely in the predicted kinematic distribution from the mock stellar sample. Here and in the following analysis, we assume a fixed heliocentric distance for all stars in PW1 of 28.7 kpc, following the measurement of \(28.7\pm0.4\kpc\) in \protect{\citet{pricewhelan19}}. For reference, at the distance of the LMC the transformation from angular to projected linear distance is \(\approx0.87\) kpc/\({}^\circ\). At the distance of PW1, the transformation is \(\approx0.5\) kpc/\({}^\circ\).

\begin{figure*} \centering \includegraphics[width=7.in]{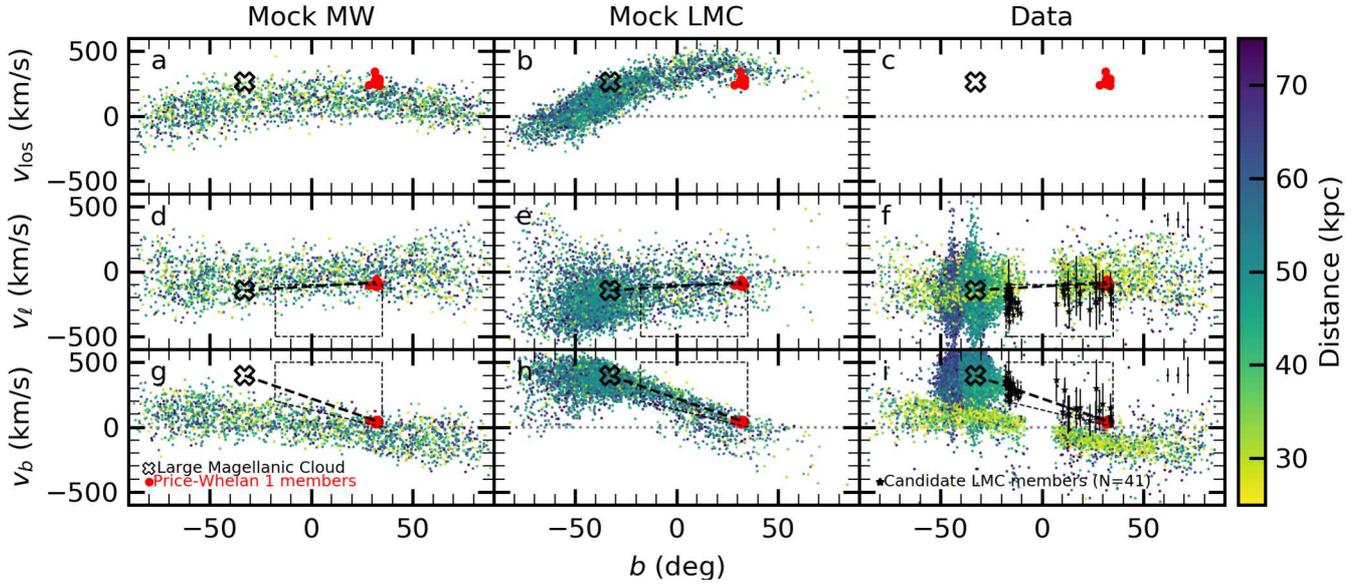} 
\caption{\label{fig:allkinematics} Comparison of mock kinematics to observed data as a function of galactic latitude \(b\). From top to bottom, we show \(v_{\rm los}\) (panels a,b,c), \(v_\ell\) (panels d,e,f), and \(v_b\) (panels g,h,i), each colour-coded by distance. In each panel, we have filtered the relevant dataset to select only particles or stars within \(30^\circ\) of the Magellanic System plane \protect{\citep{nidever08}}. From left to right, we show mock MW stars (panels a,d,g), mock LMC stars (panels b,e,h), and true RR Lyrae data (panels f,i). There are no available \(v_{\rm los}\) observations for the RR Lyrae dataset (panel c). In all panels, the `X' marks the observed location of the Large Magellanic Cloud. Red points are the stars in the proposed leading-arm star cluster Price-Whelan 1 \protect{\citep{nidever19b}}. To facilitate identification of LMC particles in the leading arm, we draw tracks between the LMC centre and Price-Whelan 1 (black dashed lines) as well as polygons outlining regions in velocity space dominated by LMC stars, which define our search space (Sec~\ref{subsec:modelfindings}). In panels f and i, we show the 20\(^{\rm th}\), 50\(^{\rm th}\), and 80\(^{\rm th}\) percentile velocity errors for the stars in the selection region. The stars at \(b>0^\circ\) in the kinematic search space have larger typical errors than stars at \(b<0^\circ\) owing to their larger average distance.} \end{figure*}

In Figure~\ref{fig:allkinematics}, we show observable kinematics for the mock stellar halos of the MW and LMC as a function of galactic latitude\footnote{We are motivated to choose \(b\) rather than \(\phi_2\) by the relatively close alignment of the the MS coordinate system and galactic latitude in the leading arm region, to avoid strong assumptions about the track of the leading arm by overinterpreting the MS coordinate system, and to facilitate inspection of additional data sets without coordinate transformations.}. We define a on-sky filtered sample for the mocks by selecting all stars satisfying three criteria: \(25<d/{\rm kpc}<75\), \(|\phi_2|<30^\circ\), and \(\ell<0^\circ\). Following the coarse model validation, we place the observable signposts on relevant panels in Figure~\ref{fig:allkinematics}. The LMC location in each panel is an un-filled `x', and the leading arm cluster PW1 is red points. Noticing that the point distribution in the clean LMC sample forms a roughly continuous curve between the location of the LMC and PW1, we select the LMC and PW1 as `anchor points' in our analysis and define a simple kinematic `track' for the LMC stellar halo as a straight line between the two. Near the trajectories, the kinematic space is dominated by LMC stars. The perpendicular velocity component \(v_b\) appears to show the strongest power to distinguish LMC stars, followed by \(v_{\rm los}\). The azimuthal component \(v_\ell\) shows little distinguishing power.

We use the above mocks to design a strategy to search for stripped halo stars in 5D coordinates \((\ell,b,d,v_\ell,v_b)\). The \citet{nidever08} coordinate system forms the first selection: we broadly consider candidates to be stars at \(|\phi_2|<30^\circ\), where \(\phi_2=0^\circ\) is the MS plane. We are interested in studying the more distant regions of the putative leading arm LMC stellar halo, so we restrict our analysis to \(b>(b_{\rm LMC}+21^\circ)\), beyond the extent probed by \citet{nidever19a}. We limit the search in the north to \(b<(b_{\rm PW1})\), both using PW1 as an anchor point for the search and also to avoid approaching the Sgr plane in the north. With the defined region in \((\ell,b)\) -- shown as the black dashed outline in each panel of Figure~\ref{fig:skydensity} -- we broadly consider stars in \(|d-d_{\rm LMC}|<25 \kpc\). The results do not change significantly if we choose 20 or 30 instead of 25 \(\kpc\). We choose this wide distance cut because the stellar halo is expected to be roughly 3d, so we want to consider depth effects as well. Additionally, the true trajectory of the LMC is uncertain, so the leading arm may occupy a large range in distances.

The kinematic cuts provide the most stringent selection criteria. From inspection of Figure~\ref{fig:allkinematics}, the MW halo follows a clear track in \(v_b\), and has effectively zero mean in \(v_\ell\). In the mock, the root variance at any given \(b\) is \(\approx 100\kms\). Using the median \(v_\ell\) and \(v_b\) as a function of \(b\), we define boundaries offset 100\kms from the median\footnote{The \(v_\ell<-100\kms\) limit is more stringent than the mean \(v_\ell\) for PW1 members, \(\langle v_\ell\rangle=-93\kms\), but PW1 is located in a region that may be difficult to disentangle from the MW halo.}. Lastly, to avoid stars with spurious measurements, we define a maximum velocity in both \(|v_\ell|\) and \(|v_b|\) of 500\kms. In order to be considered a candidate LMC member, a star must appear in {\it both} polygons. Reviewing the model predictions in \(\vlos\), it is clear that the addition of \(\vlos\) measurements would significantly help to associate RR Lyraes with the LMC. 

From this model, we conclude that if the LMC has an extended stellar halo the unique signature in filtered position and kinematic space will enable its discovery. We find that contamination rates from MW halo stars are very low in the model: \(<1\%\). In the next section, we introduce one data set with which to look for the extended LMC stellar halo.

\section{Data comparison} \label{sec:datadescription}

\subsection{RR Lyrae sample} \label{subsec:sample}

As luminous tracers with all-sky coverage from Gaia, our primary diagnostic tool is a sample of RR Lyrae stars with five-dimensional data. We follow the cleaning process from \citet{iorio19} with modifications to optimise for the recent Gaia EDR3 data release. We first obtain the Gaia DR2 {\tt vari classifier result} \citep{holl18} and specific object study \citep{clementini19} tables. We restrict our analysis to fundamental-mode pulsators to reduce ambiguity in the period-luminosity-metallicity relationship. We then select all RRab-classified stars and find the union of the two tables on Gaia DR2 ID. Next, we cross-match the resulting catalog of stars against Gaia EDR3 \citep{gaiaedr3} using 1 arcsecond tolerance. After inspecting matches, we enforce a matching source distance tolerance of 0.036 arcseconds and require that {\tt ruwe}\(<\)1.4. We compute the extinction along the line of sight {\tt dustmaps} \citep{dustmaps}, using the \citet{schlegel98} maps. We remove stars in moderate-to-high extinction regions, defined as \(A_G = 2.27E(B-V) > 0.5 \), following the calibration in \citet{iorio19}. We remove stars classified as members of the Sagittarius stream by cross-matching membership of the input RR Lyrae with the catalog of \citet{ramos20}. We remove all stars within 3 scale radii of globular cluster centres, using the catalog from \citet{vasiliev19}. We identify potential dwarf galaxy members using the positions from the compilation in \citet{mcconnachie20}. For Sextans, Sculptor, Fornax, Draco, and Ursa Minor, we remove all stars within \(1^\circ\). For all other dwarfs, we remove all stars within \(0.5^\circ\). We compute the distance using the Gaia-band extinction relation and absolute magnitude \(M_G=0.64\pm0.25\) from \citet{iorio19}\footnote{\protect{\citet{iorio21}} further demonstrates the robust nature of these measurements for DR2. The modest changes to photometry between DR2 and EDR3 are absorbed in the absolute magnitude uncertainty.}.  We assume a flat, conservative 16\% distance uncertainty. Proper motions and associated errors, including correlations, are drawn from the Gaia EDR3 catalog. We convert the proper motions into angular velocities, using \(v_{\{\ell,b\}} = 4.74\cdot d\cdot\mu_{\{\ell,b\}}\), where 4.74 is a geometric factor to convert between km/s and milliarcseconds per year. The final clean sample of RR Lyrae stars has 65723 members.
We use the cumulative heliocentric distance curve of this sample to create the mock dataset (Sec.~\ref{subsec:construction}).

In panel c of Figure~\ref{fig:skydensity}, we show the on-sky densities of RR Lyrae stars in our clean sample. Owing to incompleteness in Gaia RR Lyrae across the sky \citep{mateu20}, one must take care when interpreting (a lack of) overdensity structures. Fortunately, we benefit from the proper motion accuracy of the EDR3 data release, which enables a 5d study of the RR Lyrae sample to \(d\simeq75\) kpc. As the LMC and SMC are obvious in RR Lyraes, and have measurable and constrained proper motions, we are confident that while we are near the Gaia limit, we are not beyond for the problem at hand.

\subsection{The LMC stellar stream} \label{subsec:extendedhalo}

Applying all criteria from above, as well as the polygon selection, we find 41 RR Lyrae stars consistent with membership in the LMC stellar halo at \(\theta_{\rm LMC}>21^\circ\) (where \(\theta_{\rm LMC}\) is the on-sky angular distance to the LMC centre as defined in Sec.~\ref{subsec:initialconditions}), including 17 in the northern galactic hemisphere \(\theta_{\rm LMC}>40^\circ\), as shown in Figure~\ref{fig:allkinematics}. The mean distance of the stars in the polygon selection is 50.4~kpc, which rises to 61.6~kpc when considering only stars in the northern hemisphere. 
To address whether all of these stars are robust LMC members requires the inclusion of additional kinematic data. We come back to this issue in Section~\ref{subsec:additionaldata}.

We quantify the contribution of background stars and spurious contamination in two ways. For background (that is, the number of stars that one would expect in the MW stellar halo), we measure the stars located in an population on the opposite side of the sky: by considering stars \(|\phi_2|<30^\circ\), but in the opposite half (\(\ell>0^\circ\)) of the sky and with inverted \(v_\ell\) and \(v_b\) selection polygons. In this equivalent space, we identify \(N_{\rm background}=3\). For contamination (that is, possibly spurious members appearing in the box) we offset the velocity selection polygons to negative  \(v_b\) (positive \(v_\ell\)). In this box selection, we identify \(N_{\rm contamination}=4\). Given the candidate numbers above, the measurements of candidate LMC stars -- even in the northern hemisphere -- appear robust. We have also confirmed that the ancient merger model of \citet{naidu21} does not suggest any contamination in the selection region from theorised ancient merger halo stars. We cannot rule out other unknown substructures as possible contributors.

\begin{figure} \centering \includegraphics[width=3.4in]{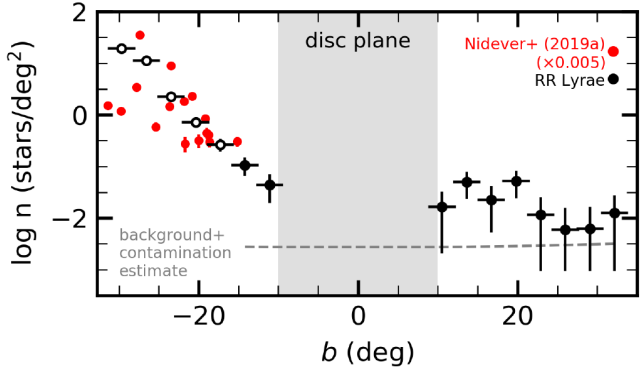} 
\caption{\label{fig:projecteddensity} The projected density of candidate LMC RR Lyrae stars (in log(stars/deg\(^2\)); black points) as a function of galactic latitude \(b\), in degrees. Uncertainties in density are estimated from Poisson statistics in each bin. Filled circles correspond to the region in the kinematic search space, while open circles correspond to RR Lyrae stars within the known stellar envelope. The grey dashed curve is the estimated contribution from background and contamination by non-LMC stars. The red points are main sequence turnoff measurements from \protect{\citet{nidever19a}}, which have been scaled by a factor of 0.005 to create approximate agreement between the samples at overlapping latitudes. We shade a region in grey where the MW disc plane makes estimates impossible. Owing to incompleteness, the estimates for star counts from RR Lyrae should be treated as {\it lower limits}.} \end{figure}

We also define a broader `known LMC space' that includes RR Lyraes within the LMC stellar envelope footprint of \citet{nidever19a} (i.e. \(\theta_{\rm LMC}<21^\circ\)). To define this kinematic space, we extend the low (high) velocity boundary in \(b-v_b\) (\(b-v_\ell\)) space to \(b=b_{\rm LMC}\). In Figure~\ref{fig:projecteddensity}, we show the density of sources, including the known LMC space, as a function galactic latitude. We caution that this density measurement should be viewed as a lower limit on the projected density owing to incompleteness. Even with such a caveat, we find that the relatively small number of RR Lyrae detected is sufficient to suggest that the stripped LMC halo extends significantly into the northern hemisphere.

\section{Discussion} \label{sec:discussion}

\subsection{Global features of the MW-LMC interaction} \label{subsec:lmcfeatures}

Our model produces key features of the MW-LMC interaction:
\begin{enumerate}
\item The reflex motion of the MW disc in response to the LMC that was predicted \citep{gomez15,erkal19,petersenpenarrubia20a} and observed \citep{petersenpenarrubia20b,erkal21,vasiliev21}. The prominent all-sky kinematic signature also confirms the displacement of the MW stellar disc from the barycentre of the total MW, an effect we readily reproduce in this model.
\item The deformation of the MW stellar halo that has been predicted \citep{garavitocamargo19,garavitocamargo21} and tentatively reported by \citet{belokurov19} and \citet{conroy21}. The deformation of the MW stellar halo can be further split into the symmetric quadrupole response excited by the LMC infall, and the asymmetric dipole feature from the displacement of the MW stellar disc from the halo. Both are observable in panel a of Figure~\ref{fig:skydensity}, with the asymmetric feature dominating near the apex, while the symmetric feature dominates the MW response in the northern sky.
\item Our model predicts the presence of LMC halo stars at angular distances \(\theta_{\rm LMC}>20^\circ\). The stellar halo exhibits a prominent leading arm, which is a telltale sign of tidal stripping. For realistic estimates of the pre-infall stellar halos of the MW and LMC, we predict that stripped material from the LMC should be visible as a stellar overdensity in the Galactic halo at distances comparable to that of the LMC (see panel b of Figure~\ref{fig:skydensity}). 
\end{enumerate}
Each effect produces unique signatures with complex patterns across tens of degrees on the sky. Kinematics and accurate models are required to fully disentangle them. For example, \citet{conroy21} report the detection of the LMC wake in a regions of the Northern hemisphere that overlap with the stripped LMC halo, possibly confusing the signals. Kinematics are key to separate the two features. Interestingly, we find that the selections on distance and proper motion cuts of \citet{conroy21} efficiently remove most LMC stream members. Applying our kinematic search parameters to the \citet{conroy21} sample of giant stars, we find that two of the stars (out of 1301) may be members of the LMC stream. These stars are located \(\theta_{\rm LMC}=\)18.4\(^\circ\) and 24.7\(^\circ\). 

Recent studies of the LMC have found evidence for disc structures out to \(\theta_{\rm LMC}=15^\circ\) \citep{cullinane20,gaia_lmc21}. Other efforts to map the outskirts of the LMC using main sequence turnoff stars have found evidence for a stellar halo \citep{majewski09,nidever19a}, but only out to modest distances. \citet{nidever19a} in particular fit an exponential+power law profile with a break at \(\theta_{\rm LMC}=13-15^\circ\). Material outside this break is assumed to be part of a stellar `envelope'. Further recent work has found evidence for other extended LMC material, including substructure \citep{belokurov16,navarrete19}. 
In this study, we believe we have a bona fide detection of a LMC stellar halo populated by old stars that extends out to a projected distance of \(\theta_{\rm LMC}\approx30^\circ\) from the LMC centre. These stars show kinematics consistent with a tidally-perturbed stellar halo rather than material kicked out of the LMC disc \citep{weinberg00}, but further models are needed to study this possibility.

\subsection{Future data} \label{subsec:additionaldata}

The single most important additional dataset would be one that includes all six phase-space dimensions, as identification of tidal stream members in the MW halo is particularly straightforward in angular momentum space \citep[e.g.][]{pp21}. Even with sparse coverage, a detection of the leading arm would provide important constraints on the trajectory of the LMC, which is strongly dependent on the mass distribution of both the MW and LMC. Unfortunately, kinematic surveys of the Northern Galactic hemisphere only cover a fraction of the MW halo, and thus few line-of-sight velocities exist for rare objects that may trace the LMC stream. Additionally, \(\vlos\) is challenging to measure for RR Lyrae stars owing to their pulsation properties. However, one may be able to constrain the \(\vlos\) values in sufficiently inexpensive observations to determine if the search space is valid, and then use samples to search for LMC stellar halo members. Future data will provide more insight into the LMC leading arm and possible extended halo. We emphasise that our goal is not to exhaustively characterise the LMC stellar halo -- a difficult goal given present data limitations -- but to help define the prediction and search space for follow-up work. In particular, the stellar halo can be constrained by (i) collection of additional 5 or 6-d datasets from existing surveys, (ii) spectroscopy of the RR Lyraes for radial velocities to validate membership, or (iii) future RR Lyrae releases. 

\subsection{Future models} \label{subsec:futurework}

One $N$-body realisation is not sufficient to model the observed MW-LMC interaction. We stress that our model is not tuned to fit the outskirts of the LMC. Instead, we use a live $N$-body realisation to guide our search for the stripped LMC halo, in the hope that a positive detection of the leading arm of the LMC stream will facilitate a deeper understanding of the mass distribution in the LMC and the MW.

There are several relevant shortcomings in the theoretical model presented in this paper. We have effectively treated the LMC stellar halo as a scaled-down MW stellar halo, which may not be borne out. We expect that different parameter choices will likely affect the quantitative aspects of the predictions made in this paper, but will not alter our conclusions at a qualitative level. E.g. when applying the same kinematic selection region to the $N$-body model we find 245 stellar particles that belong to the LMC (versus 41 stars in the data set). A fraction of the stellar particles may be located along sight lines with high extinction, which complicates comparison against the data. Interestingly, we find that the mean distance of the leading arm in our mock is \(\approx50\kpc\), nearly independent of \(b\), whereas LMC halo candidates in the north are located at systematically larger distances. A closer inspection of the model reveals a large heliocentric distance spread in the stellar halo of the LMC, which can be traced back to the choice of the initial profile. It is also unclear whether the trajectory of the LMC is correctly reproduced by our $N$-body realisation, which would have a clear impact on the curvature and the kinematics of the leading arm in the Northern hemisphere. In particular, determining the apocentre of the leading arm of the MS would provide important constraints on MW-LMC models. It is also possible that the stellar halo trajectory in the north does not follow the canonical \citet{nidever08} track. We notice that in our mocks material stripped from the LMC departs from the great circle defined by the Magellanic System. Additional constraints on the trajectory of the LMC across the MW will be crucial to build follow-up models for the stellar stream of the LMC.

Lastly, we have not included the Small Magellanic Cloud (SMC) in our analysis. How do the signatures change when the SMC is included? As the trajectory and kinematics are largely set by the infall, the kinematic difference for the leading arm is likely to be relatively small when the SMC is included -- supported by the location of leading arm gas. However, the role of the SMC in shaping the pre-infall LMC stellar halo remains to be explored. Can repeated SMC encounters process the stellar halo and produce different velocity structure? Does the presence of the SMC imply that the LMC halo is under-luminous with respect to \(\Lambda\)CDM predictions?

\section{Conclusion} \label{sec:conclusion}

The LMC may be currently losing its stellar and dark matter halos to the MW tidal field. Here, we use a live $N$-body simulation that follows the MW-LMC interaction in order to predict the distribution and kinematics of LMC halo stars under reasonable assumptions. In our model, most halo stars are still found in the vicinity of the LMC. However, a considerable fraction of the original halo has been tidally stripped and populates a prominent stellar stream that extends into the Northern Galactic hemisphere.

Our $N$-body model predicts a clear signature of the leading arm at the distance of the LMC. In particular, stripped halo stars exhibit both high \(v_b\) and \(v_{\rm los}\) when compared to the MW stellar halo in the same region on the sky. We used a sample of Gaia RR Lyrae stars to search for the predicted kinematic signatures. We find solid evidence for halo members at moderate (though larger than reported in the literature to-date) distances \((\theta_{\rm LMC}=20-30^\circ)\) from the LMC centre, and tentative evidence for stripped RR Lyrae stars associated with the leading arm in the Galactic north.

While far from conclusive, the comparison of a mock dataset drawn from a realistic MW-LMC model and the newly-released Gaia EDR3 proper motions suggest that there may be a large sample of LMC stellar halo stars waiting to be discovered beyond the handful detected here. Measuring the extended stellar halo of the LMC will place constraints on the pre-infall assembly history of the LMC, as well as on the infalling trajectory of the LMC and the outer mass distribution of the MW.



\section*{Acknowledgements}

It is a pleasure to thank Sergey Koposov, Sophia Lilleengen, Denis Erkal, and Nico Garavito-Camargo for comments and discussions on early versions of this work. MSP thanks Martin Weinberg for use of the {\sc exp} code and acknowledges funding from a UK Science and Technology Facilities Council (STFC) Consolidated Grant. This work used cuillin, the Intitute for Astronomy's computing cluster (http://cuillin.roe.ac.uk), partially funded by the STFC. We thank Eric Tittley for continued smooth operations. This project made use of {\it numpy} \citep{numpy} and {\it matplotlib} \citep{matplotlib}. 

\bibliography{LMCNorth}

\label{lastpage}

\end{document}